\documentclass[12pt]{article}

\usepackage{graphicx}
\usepackage{amsmath}
\usepackage{bm}
\usepackage{hyperref}
\usepackage{xcolor}
\usepackage{bbold}

\begin{document}

\title{The futility of being selfish in vaccine distribution}

\author{
  \href{https://orcid.org/0000-0001-7310-3252}{\includegraphics[scale=0.06]{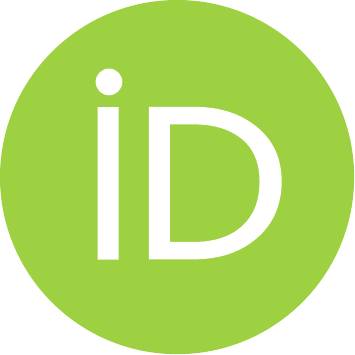}\hspace{1mm}Felippe Alves}, \href{https://orcid.org/0000-0001-9821-2623}{\includegraphics[scale=0.06]{orcid.pdf}\hspace{1mm}David Saad} \\
  Department of Mathematics\\
  Aston University \\
  Birmingham, UK\\
  \texttt{alvespef@aston.ac.uk, d.saad@aston.ac.uk} \\
}



\date{\today}



\maketitle

\begin{abstract}
We study vaccine budget-sharing strategies in the SIR (Susceptible-Infected-Recovered) model given a structured community network to investigate the benefit of sharing vaccine across communities.
The network studied comprises two communities, one of which controls vaccine budget and may share it with the other. Different scenarios are considered regarding the connectivity between communities, infection rates and the unvaccinated fraction of the population. Properties of the SIR model facilitates the use of Dynamic Message Passing (DMP) and optimal control methods to investigate preventive and reactive budget-sharing scenarios. Our results show a large set of budget-sharing strategies in which the sharing community benefits from the reduced global infection rates with no detrimental impact on its local infection rate.
\end{abstract}

\maketitle


\section{\label{sec:intro} Introduction}

Epidemic spreading processes give rise to various challenges, from effective policy making for containment and mitigation, formulating balanced isolation rules and preventing a healthcare system overload; a failure in dealing with the epidemic may lead to excess death and suffering in the population, and a breakdown of the health provision system. Part of the problem is the inability to reach sufficient worldwide vaccine coverage and prevent the appearance of new variants. 
In response to the COVID-19 pandemic, we have had the fastest response in history in developing vaccines for a new virus, exploring different techniques and achieving multiple immunization alternatives in less than a year.
However, as vaccine production became a reality, so did the disparity in its distribution across countries.

According to Data Futures Platform~\cite{ExploreData}, some of the richest countries may have, at the time of writing, more than $3$ times the number of doses necessary to fully immunise their entire population against COVID-19, while poorer countries do not sufficient vaccines for a full coverage~\cite{ExploreData}.
A recent study~\cite{mooreRetrospectivelyModelingEffects2022} based on this data, estimates retrospectively the number of lives that could have been saves with a better vaccine distribution strategy. Relying on an extended version of the mean-field SIR (Susceptible-Infected-Recovered) model, that considers the estimated infection ratio, the appearance of new variants and details of the strategies employed in the effort to limit the spread of COVID-19; a different work shows that distributing the vaccine accordingly to the non vaccinated population would have a significantly reduced total number of deaths~\cite{ledfordCOVIDVaccineHoarding2022}.

Spreading processes are ubiquitous in social, natural and technological networks. They play an increasingly important role in opinion setting, marketing and epidemic modeling~\cite{anderson1992infectious,boccaletti2006complex,rogers2014diffusion,pastor2015epidemic}. 
While cascading effects in spreading processes can be desired from an informative or persuasion campaign perspective, such as election campaigns~\cite{rutledge2013obama,epstein2015search,margetts2015political,domingos2001mining} or for raising public awareness~\cite{IceBucketChallenge}, being able to predict and control cascading effects becomes essential to prevent economic loss and unnecessary deaths~\cite{ferc2012arizona,dawood2012estimated,act2013financial,lokhov2016detection}.
The key to understanding the dynamics of spreading processes lies in how information, viruses or failures flow through the edges of an interaction network between individual constituents, allowing for the identification of important nodes in the spreading processes and other topological features where control can be employed to contain or boost the spreading process.
On that front, there are multiple strategies for optimal resource allocation in different spreading scenarios. 
One approach, based on topological measures, focuses on identifying influential spreaders using various topological measures, such as high-degree nodes, betweenness centrality, random-walk and graph partitioning, among other~\cite{pastor2002immunization,cohen2003efficient,holme2002attack,holme2004efficient,chen2008finding,kitsak2010identification}.
A caveat of algorithms based on topological properties is their variable performance depending on network instances and process dynamics~\cite{borge2012absence,hebert2013global}.

The problem of finding an optimal immunisation set has been addressed using belief propagation~\cite{altarelli2013optimizing,altarelli2014containing,guggiola2015minimal,kempe2003maximizing,chen2013information,Lokhov_2017,liImpactPresymptomaticTransmission2021a} for the SIR, SEIR (Susceptible-Exposed-Infected-Susceptible) and SIS (Susceptible-Infected-Susceptible) models.  

In this study we investigate the SIR model in a network with known community structure to investigate the possibility of sharing vaccine doses across communities (e.g. countries) being beneficial to communities with surplus doses. It is important to point our that the main thrust of the paper is neither to model a specific realistic epidemic spreading nor to address the very intricate distributions and parameters involved in realistic modelling. The aim is to show that sharing vaccines across communities would benefit both communities due to the cross links between them.

We set up a simple network with two communities, one of which controlling a vaccine budget and 
shares it with the other.
The simplicity of the SIR model allow us to employ Dynamic Message Passing (DMP)~\cite{lokhovDynamicMessagepassingEquations2015a} for estimating the marginal distribution of each node state.Combining DMP and Optimal Control methods, we analyse possible approaches to manage a vaccine budget to preemptively mitigate the effects or fight an ongoing epidemic of possible new variants.
We also account for a few different scenarios regarding the connectivity between each community and the fraction of population that cannot or refuse to be vaccinated.
\begin{figure}
    \centering
    \includegraphics[width=0.75\linewidth]{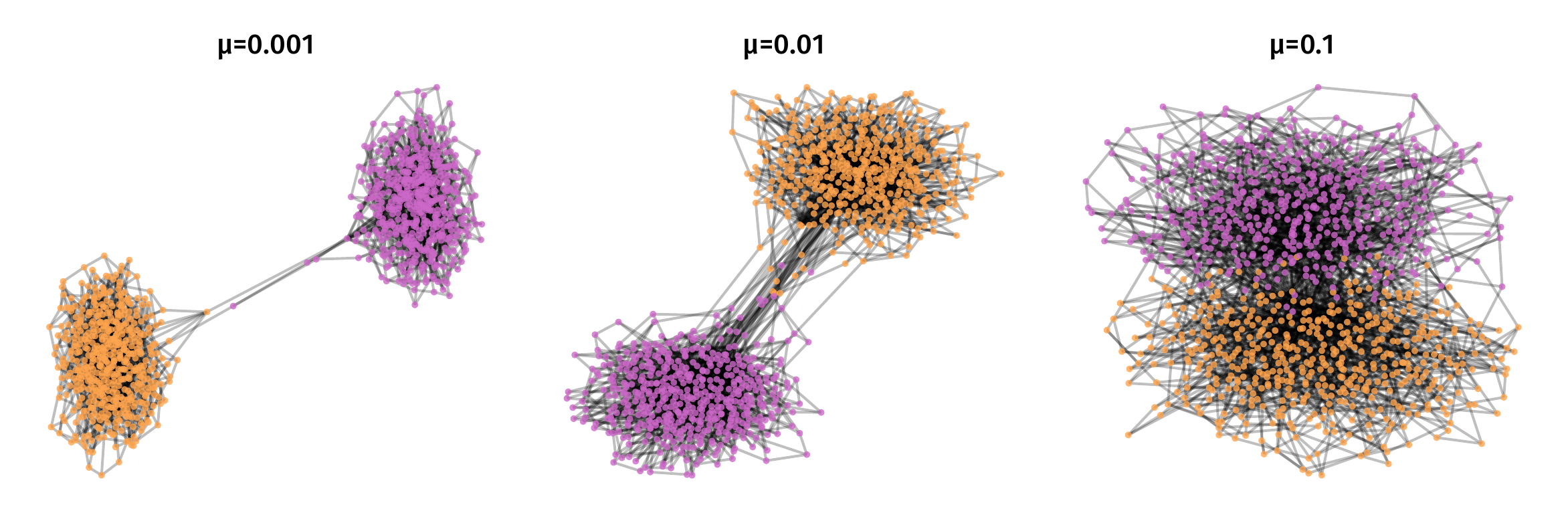}
    \caption{Example networks with node degree given by a power law distribution with degree $\alpha=3.1$, minimum node degree $k_0=3$ and a fraction $\mu \in \{10^{-3},10^{-2},10^{-1}\}$ of edges between communities.}
    \label{fig:example-network}
\end{figure}
Our results show a large set of sharing strategies for which the community controlling the vaccines pays little to no cost in terms of its infected population while being globally beneficial.

In Sec.\ref{sec:theory} we describe the SIR model in one of its simplest variations, in Sec.~\ref{sec:DMP} we introduce the Dynamic Message Passing (DMP) method for analysing it, the choices for network topology. A discussion of the results follows in Sec.~\ref{sec:discussion}.
The source code with the implementation of the algorithms, analysis and figures can be found at~\cite{codegtilab}. 

\section{\label{sec:theory}Model}
While a dynamic contact network can be easily accommodated within the same framework, the spreading process studied here takes place in a static contact network, represented by a graph $\mathcal{G}=(\mathcal{V},\mathcal{E}) $ with nodes $\mathcal{V}$ representing individuals and edges $\mathcal{E}\in \mathcal{V}\times \mathcal{V}$ representing interactions between them. Each node $i$ is in a state $\sigma_i\in \{S,I,R\}$ indicating whether it is susceptible, infected or recovered (or vaccinated). The epidemic evolution is encoded by the state transition rules

\begin{equation} \label{eq:sir-rules}
\begin{split}
    \beta^{ij}: S^i + I^j & \to I^i + I^j \\ 
    \gamma^j: I^i & \to R^i~.
\end{split}
\end{equation}
Transition between states, from susceptible to infected is due to having an infected neighbor is controlled by the parameter $\beta^{ij}$, the probability that the infected node $i$ infects the neighboring susceptible node $j$; the coefficient $\gamma_i$ represents the probability of spontaneous recovery of the infected node $i$. 

The transition rules~\eqref{eq:sir-rules} describe the interaction between nodes at each time step $t\in \{0,\dots,T\}$, which may correspond to days. At each time step an infected node $i$ recovers with probability $\gamma^i$ and, therefore, the average recovery period is $1/\gamma^i$. Analogously, the average period before a susceptible node $i$ is infected by any infected node is $1/\sum_{j\in\partial i}\beta^{ji}$. \emph{For demonstration purposes only}, we use the values for recovery and incubation periods for COVID-19; we set $\gamma^i = \gamma = 1/8$ and $\beta^{ij} = \beta = 1/5$, respectively~\cite{liImpactPresymptomaticTransmission2021a}. The values of $\gamma^i=\gamma$ and $\beta^{ij}=\beta$ are assumed uniform across nodes and edges and chosen so that the SIR dynamics is at equilibrium at $T$. Variable parameters between sites and in time can be easily accommodated within the same framework.

In this study, we investigate the conditions for a vaccine-rich community to benefit from sharing it with a neighboring vaccine-poor community with no access to it.
To investigate how sharing can affect the spreading process, we set up a network with two communities, $A$ and $B$, connected by a randomly chosen fraction $\mu$ of the total number of edges, which translates to a fraction $\mu$ of nodes in $A$ connected to $B$. 

\section{\label{sec:DMP}Analysis - Dynamic Message Passing}
While the infection dynamics can be simulated computationally from random initial conditions, obtaining reliable solutions is computationally demanding as the system size grows. Alternative methods that assumes uniform spread~\cite{volz2008sir} or connectivity~\cite{li2010epidemic} are also inaccurate when one deals with a specific instance of the graph. Exact solutions of the spreading process defined by Eqs~\ref{eq:sir-rules} are hard to obtain and various probabilistic approximation methods have been developed to tackle this type of complex dynamics, such as individual-based mean-field, belief propagation~\cite{Altarelli2014} and dynamic message passing (DMP). An overview of the features of the former how dynamic message passing can address some its issues are given in~\cite{liImpactPresymptomaticTransmission2021a}.
In this work we employ the DMP equations for the SIR model derived in~\cite{lokhovDynamicMessagepassingEquations2015a}.

The DMP algorithm estimates the marginal distribution of node-states by iterating local message exchanges in the form of conditional probabilities. As most message-passing based algorithms it is exact on trees but offers a good approximation for locally tree-like networks. For the SIR model, in which the interest lies in estimating the marginal distribution for each node, it is possible to track the marginal estimates by introducing a set of messages for each directed edge and a set of messages for each node. The evolution of each message can be deduced through probabilistic inference using the spreading process rules~\eqref{eq:sir-rules}. At any step $t$, a node is susceptible with probability $P^i_S(t)$ given by the probability $P^i_S(0)$ of being susceptible at $t=0$ and the probability is has not been vaccinated or infected up to time step $t$. A susceptible node is vaccinated at $t$ with probability $\nu^i(t)$ and the probability it has not been infected by neighbor $k$ up to time $t$ is $\theta^{k\to i}(t)$, leading to $P^i_S(t) = P^i_S(0)\prod_{\tau=0}^{t-1}(1-\nu^i(\tau))\prod_{k\in\partial i}\theta^{k\to i}(t)$. 
Node $i$ can become recovered at $t+1$ is with probability $P^i_R(t+1) = P^i_R(t) + \nu^i(t)P^i_S(t) + \gamma P^i_I(t)$, considering the recovery rate $\gamma$; the probability of $i$ being infected is the complement of not being infected $P^i_I(t) = 1 - P^i_S(t) - P^i_R(t)$.

The evolution of the message $\theta^{k\to i}$ requires another edge message $\phi^{k\to i}(t)$ representing the probability of a neighbor $k$ being infected but not infecting node $i$ up to time step $t$. The message $\theta^{k\to i}$ decreases as $\phi^{k\to i}$ increases, since having an infected neighboring state increases the probability of spreading the infection, leading to $\theta^{k\to i}(t+1) = \theta^{k\to i}(t) - \beta\phi^{k\to i}(t)$.
The message $\phi^{k\to i}$ is reduced with the probability of the neighbor $k$ not being infected at $t$ for similar reasons. The probability, $\phi^{k\to i}$, is given by the probability of neither activating 
the infection signal $\beta$
nor the recovery $\gamma$, or by increasing the probability of being susceptible in the cavity graph where $i$ is removed (hence the name, the graph with a cavity where node $i$ has been). The later is given by a another edge message $P^{k\to i}_S$. In analogy with the dynamics of $P^i_S(t)$, its evolution is given by the probability of $k$ being susceptible at step $t=0$ and not being infected by neighbors in the cavity graph without $i$, i.e. $P^{k\to i}_S(t) = P^k(0)\prod_{\tau=0}^{t-1}(1-\nu^k(\tau))\prod_{j\in\partial k/i}\theta^{j\to k}(t)$. Therefore, the evolution of $\phi^{k\to i}$ is given by $\phi^{k\to i}(t+1) = (1-\beta)(1-\gamma)\phi^{k\to i}(t) - [(1-\nu^k(t+1))P^{k\to i}_S(t+1) - (1-\nu^k(t))P^{k\to i}_S(t)]$.
\section{\label{sec:results}Results}
In this section we will review a number of spreading and optimization objective scenarios using the framework described in Sec.{\ref{sec:theory}}.

\subsection{Sharing vaccines to fight an ongoing epidemic}

In the first instance, our main objective is to minimise the expected number of infected nodes within community \emph{B} throughout the duration of the epidemic, which is given by $\mathcal{I}_B = \mathbb{E}\left[\sum_{i\in B}\sum_{t=1}^{T-1}\mathbb{1}[\sigma_i^t = I]\right] = \sum_{i\in B}\sum_{t=1}^{T-1} P_I^i(t)$. However, other objectives can be chosen, for instance, flattening the infection curve to prevent overloading the health system. Introducing entropy-like non-linearity will give rise to a vaccination strategy that result in a more uniform infection over time. Minimising the expected number of infected nodes in \emph{B} in conjunction to having a more uniform infection spread, can be achieved by maximizing the objective function (minimizing entropy)

\begin{align}
    \mathcal{O} & = \sum_{i\in V}\sum_{t=1}^{T-1}\mathbb{1}[i \in B](1 - P_I^i(t))\ln(1-P_I^i(t)) \nonumber\\
    & = \sum_{i \in B}\sum_{t=1}^{T-1}{(P_S^i(t)+P_R^i(t))\ln{(P_S^i(t)+P_R^i(t))}}
\end{align}

subject to the constraints enforcing the DMP equations, initial conditions and vaccine budget. The particular choice for the $p\ln p$ non-linearity was made to favor the Susceptible state, since vaccination and Recovery are conflated and allowing nodes to be infect and to recover could lead to deceptively higher objective function values. 

The optimization employs methods from DMP and optimal control~\cite{Lokhov_2017} to maximise the objective using forward-backward dynamics, where the DMP forward dynamics is complemented by a backward optimization dynamics of the Lagrange multiplier variables.  
Maximising the objective function allows us to choose an optimized vaccination strategy $\nu^i(t)$ for every node and all times. See appendix~\ref{appendix-dmp} for the details.
To manage a budget $N V(t)$ of vaccines, the vaccination probabilities $\nu^i(t)$ must satisfy $\sum_{i\in \mathcal{V}}\nu^i(t) = N V(t)$ at all times, where $V(t) \in [0,1]$ represents the fraction of the society that can be vaccinated at $t$.

In reality, different scenarios of vaccine availability may occur, considering the difficulties involved in development, production and distribution. While different scenarios can be accommodated by the current framework, to simplify our analysis and focus on the major problem of vaccine shortage, we consider scenarios where throughout the time window considered, a total of $N V$ nodes will be vaccinated, uniformly through time, i.e. $V(t) = V/T$. We assume that community $B$ has control over a budget of vaccine doses $V$, represented as a fraction of nodes to be immunized in the whole network and chooses how to share this budget with community $A$, with a split $\rho\in[0,1]$.

\begin{figure}[!ht]
    \centering
    \includegraphics[width=0.65\linewidth]{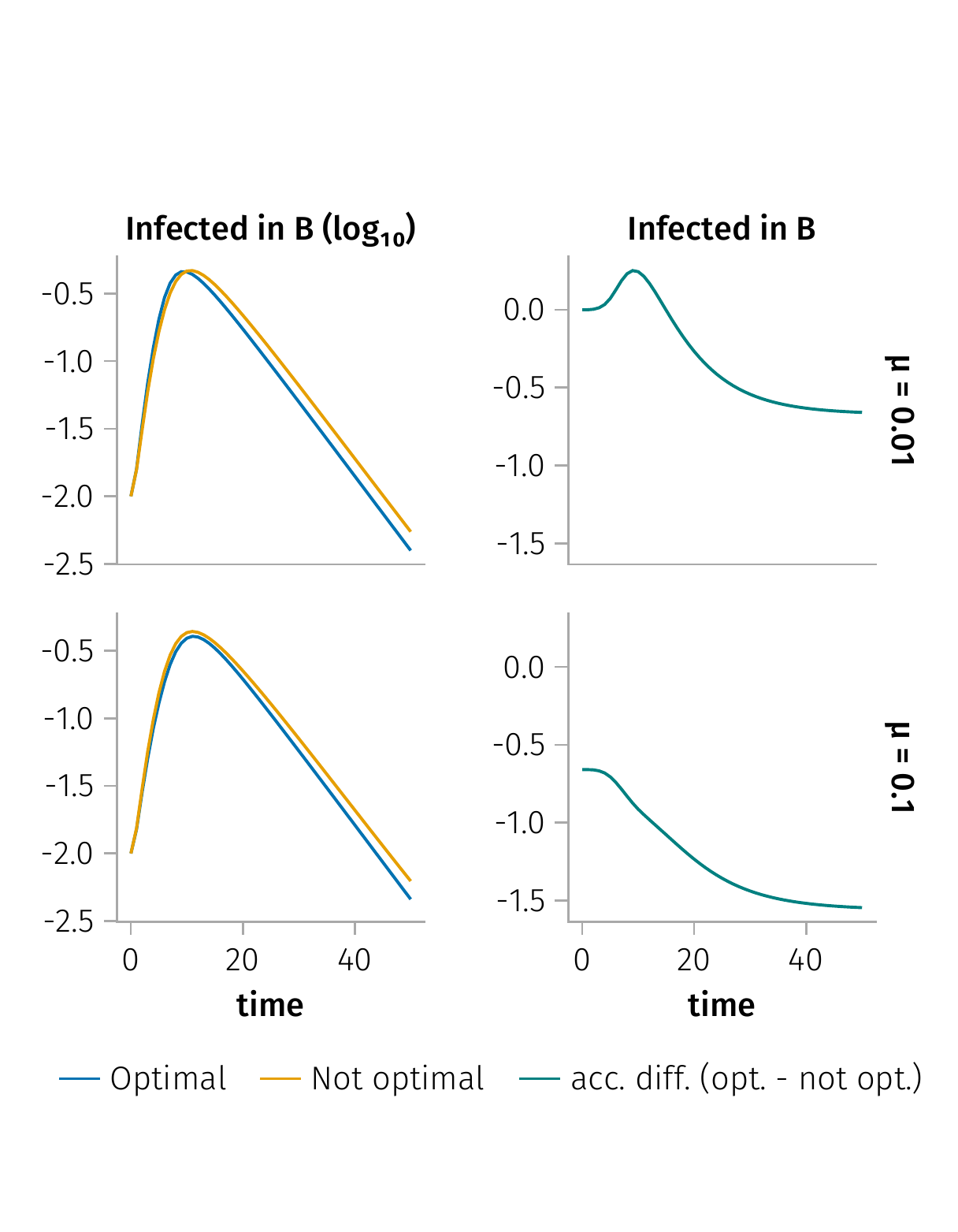}
    \caption{The optimal vaccination strategy with a budget of $V = 0.5 N$ and two connectivity scenarios: where $\mu=0.01$ (top figures) and $\mu=0.1$ (bottom figures) of the total edges connecting communities $A$ and $B$, respectively. Figures on the left column show the evolution of fraction of infected nodes in $B$ with and without optimal sharing; on the right is shown the accumulated difference in infected nodes between the optimal split and not optimized vaccination, i.e. $\sum_{t'=0}^t I_B^{\mathrm{opt}}(t') - I_B(t')$. In both cases it is clear that community $B$ is better off vaccinating its own constituents and increasingly shares the vaccines with community $A$.}
    \label{fig:optcon-split-evo}
\end{figure}

\begin{figure}[!ht]
    \centering
    \includegraphics[width=0.65\linewidth]{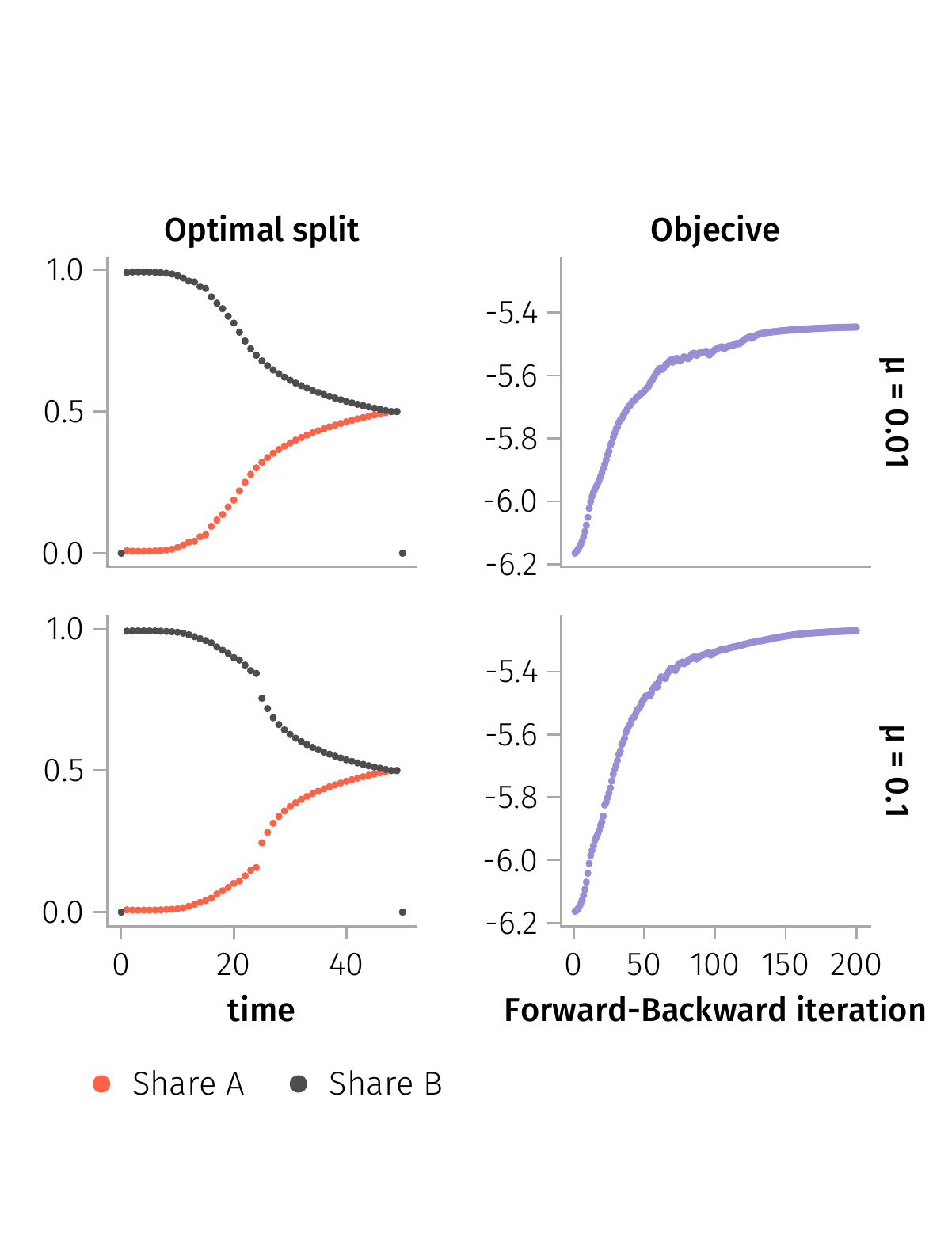}
    \caption{The optimal vaccination strategy with a budget of $V = 0.5 N$ and two connectivity scenarios: where $\mu=0.01$ (top figures) and $\mu=0.1$ (bottom figures) of the total edges connecting communities $A$ and $B$, respectively. The left column shows the optimal sharing probabilities; the right is the objective value optimization under forward-backward optimal control cycles.}
    \label{fig:optcon-split-evo2}
\end{figure}

Following a similar approach to~\cite{lokhov2014inferring,Lokhov_2017}, we use the optimal control method jointly with a forward-backward update algorithm to optimise the objective $\mathcal{O}$ and obtain a optimal vaccination strategy $\nu^i(t)$. Of particular interest are cases where assigning a vaccination probability to nodes in $A$ reduce the number of infected nodes in $B$.
Figure~\ref{fig:optcon-split-evo} shows that optimizing the distribution is beneficial to $B$ and figure~\ref{fig:optcon-split-evo2} shows that the most beneficial strategy for $B$ is to initially focus on vaccinating its own nodes and eventually share its budget with $A$.

\begin{figure}
    \centering
    \includegraphics[width=0.75\linewidth]{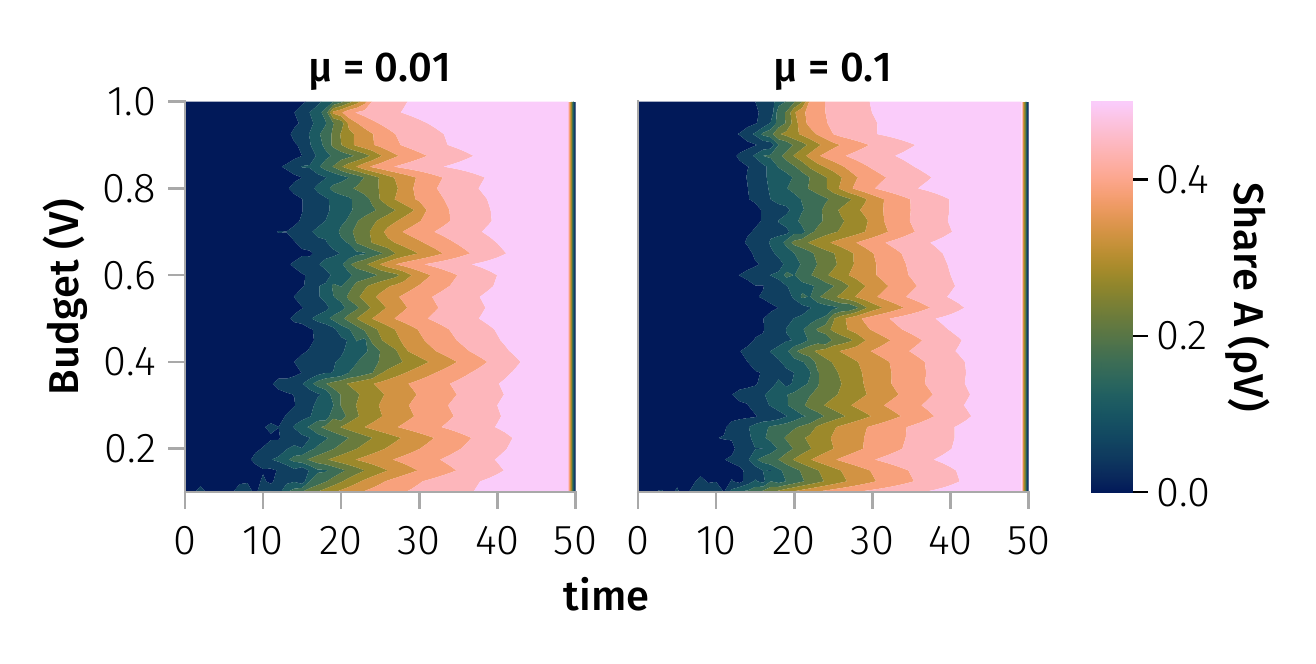}
    \caption{The best budget split at all times for budget values $V \in [0.1,1.0]$ in two connectivity cases with $\mu=0.01$ and $\mu=0.01$ of total constituents of $A$ and $B$ being connected. Sharing starts later for more highly connected communities.}
    \label{fig:optcon-split-hm}
\end{figure}

Although the optimal split varies, the general tactic of $B$ is first to vaccinate the constituents of $B$ and then share share the vaccines budget with $A$ to reduce the number of infections in $B$. Figure~\ref{fig:optcon-split-hm} shows the share of the budget that $B$ should give to $A$ at each time for two connectivity scenarios. It seems that sharing starts later for more highly connected communities.

\subsection{Preemptive vaccine sharing} 
To investigate preemptive sharing, we set $\nu^i(t)=0$ and choose uniformly the initial probability of each node to be recovered proportional to the budget available for each community. We also acknowledge the inability to immunize every node, representing anti-vaccine or vaccine-sensitive population, encoded as a fraction $x$ of nodes that cannot be vaccinated at $t=0$. 
Finally, we focus on how sharing the vaccine can mitigate the impact of new variants, assuming that exactly one node infected at $t=0$.
The objective, from the perspective of community $B$, is to reduce the impact of the disease on its nodes.
Multiple measures of impact on $B$ can be considered, such as the total number of infected nodes at any time ($\frac{1}{T}\sum_{i\in B}\sum_{t=1}^{T} P_I^i(t)$), the peak of infection ($\operatorname{argmax}_t \sum_{i\in B} P_I^i(t)$) and the influx infections coming from contact with constituents of $A$.
A proper strategy for mitigation would require one to devise a vaccine distribution strategy to maximize the benefit to its own community, potentially giving the proper weight to each impact measure. The current framework is very flexible and can accommodate each of these measures.

A vaccine budget $V\in[0,1]$ representing the fraction of nodes that can be immunized at $t=0$, is split between communities $A$ and $B$ assuming the later has control over the share ratio $\rho$. It is also assumed that a single node is infected at $t=0$, representing the appearance of a new variant and that the vaccine is still effective against it.
We also take into consideration the impossibility of vaccinating all nodes for reasons such as vaccine deniers, people who do not care or that do not engage with the health systems, by considering a fraction $x$ of nodes that cannot be immunized. 

In terms of initial message values, an infected node among the total of $N = N_A + N_B$ nodes leads to  a probability of infected nodes $P^i_I(0)=1/N$ for all nodes $i$ in the network, while 
$P^a_R(0)=\mathrm{min}\{(1-x)\rho V N / N_A,1\}$ for nodes $a$ in community $A$ and $P^b_R(0)=\mathrm{min}\{(1-x)(1-\rho)V N / N_B,1\}$ for nodes $b$ in $B$, where $N_A$ and $N_B$ are the number of nodes in each community. 
Notice the possibility of excess vaccine, if community $B$ holds a budget greater than the number nodes that can be immunized in community $B$; community $B$ may decide to accumulate the excess vaccine for future use instead of sharing it, or if the share ratio is too favorable towards $A$.
For the edge messages, $\theta^{k\to i}(0)=1$, $\phi^{k\to i}(0)=P^k_I(0)$ and $P^{k\to i}_S(0)=P^k_S(0)$ for all edges $(k,i)$.

The values chosen for the parameters are given in table~\ref{tab:parameters} and the initial conditions are given in~\eqref{eq:initial-values}, both are in appendix~\ref{appendix-dmp}. Initial message values are summarized in~\eqref{eq:initial-values}, where $i$ and $k$ refer to any node in the network, while $a$ and $b$ refer to nodes in community $A$ and $B$, respectively.
%

\begin{figure}
	\centering
    \includegraphics[width=0.75\linewidth]{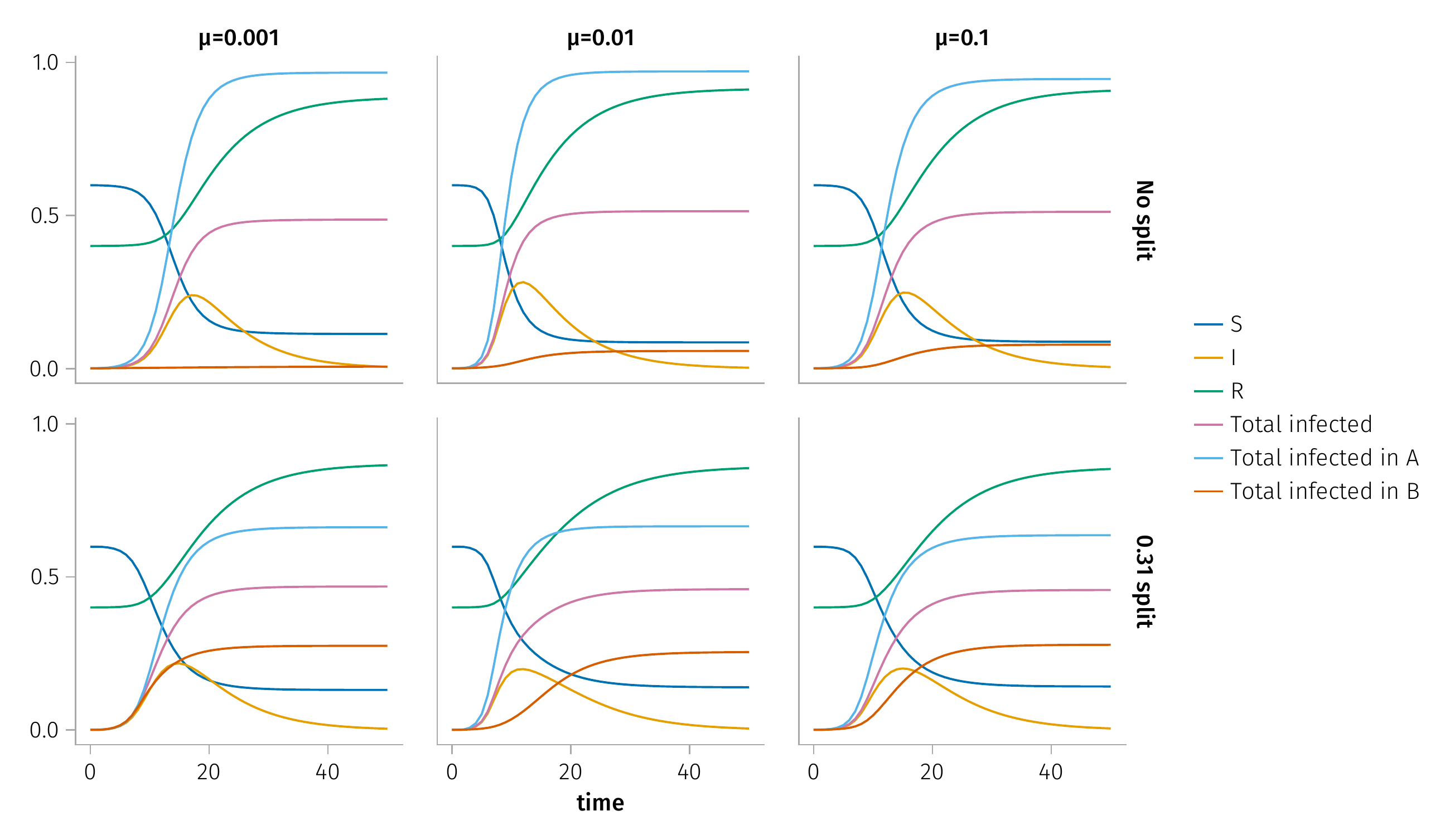}
	\caption{Epidemic evolution with $x=0.01$ fraction of excluded nodes. The total number of infected nodes is given by $1 - P_S(t) - P_R(t=0)$, where $P_R(t=0)$ is the fraction of nodes initially vaccinated; in this case, the vaccine budget of $V=0.5$ of nodes in the whole network, is enough to cover the whole community $B$. Notice how sharing can have a positive outcome globally. Each column shows the effect of a fraction $x$ of nodes in $B$ being excluded from vaccination.}
	\label{fig:evoex}
\end{figure}

The spreading process can be tracked by estimating the marginal probabilities for each state, which is the typical objective for the SIR model. 
Figure~\ref{fig:evoex} shows two examples for the evolution of the disease spreading: the first, a typical global evolution with community $A$ not having any vaccine and $B$ completely immunised (top) and another where $B$ contributing $\rho=0.25$ of its budget to $A$ (bottom).
We track all the states, globally, and the actual number of nodes that are not susceptible or initially vaccinated.

\begin{figure}
    \centering
    \includegraphics[width=0.75\linewidth]{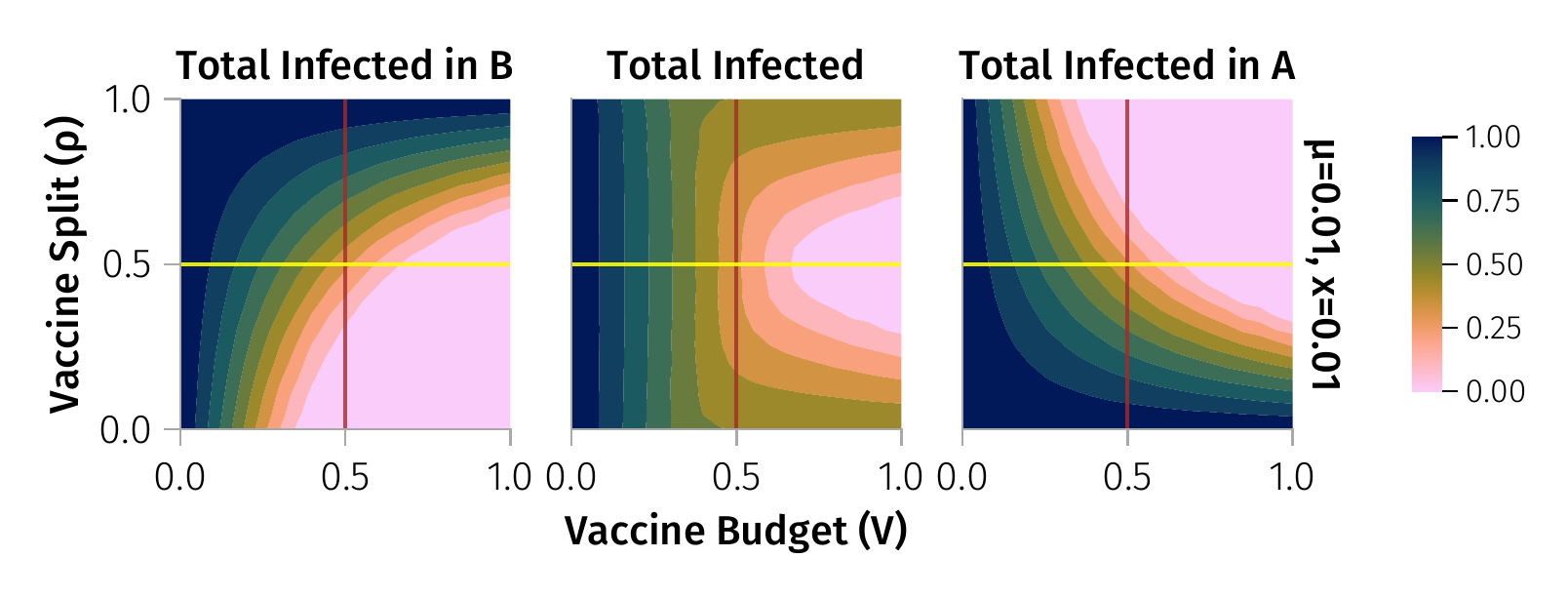}
    \caption{Total number of non-susceptible nodes for all possible budget and share values at $t=50$ and $\mu=0.001$. The brown line shows the budget necessary to immunize all nodes in $B$ yellow line shows an equal split based on community size. Notice the region in light pink shows the possibility of having a very small effect on $B$ by sharing, even in situations where full (effective) vaccine coverage would be sacrificed.}
    \label{fig:infec}
\end{figure}

Even if sharing is globally beneficial, the question whether it is futile for community $B$ to be selfish can be answered by considering in which conditions the number of infected nodes in $B$ does not increase by sharing vaccine resource.
Looking at the total number of non-susceptible nodes in $B$ as a function of the vaccine budget $V$ and share ratio $\rho$ as in Fig.~\ref{fig:infec} we realize that there is a parameter region where sharing is viable, in the sense that community $B$ is not worse off in spite of sharing their vaccines with community $A$.

A more illustrative metric for the effect of sharing can be calculated by comparing the increase in the number of infected nodes for a given share ratio against the case in which community $B$ keeps all the resource.
Let $n^c_t(\rho) = 1 - p(S|t,\rho,c) - p(R|t=0,\rho,c)$ be the fraction of nodes in states $I$ or $R$ in community $c\in\{A,B,A+B\}$ at step $t$ and for a share ratio $\rho$, excluding the nodes immunized at $t=0$.
We define the cost of share ratio $\rho$  for community $c$ as
\begin{align}
    L^c_t(\rho) & = n^c_t(\rho)\ln\frac{n^c_t(\rho)}{n^c_t(\rho=0)} \label{eq:def-cost}
\end{align}
which measures the fraction of nodes in states $I$ or $R$ (excluding vaccinated) for a share ratio $\rho$ relative to the corresponding fraction without sharing ($\rho=0$).
A negative value for $L^c_t(\rho)$ indicates that sharing vaccines  {\em reduces} the total fraction of infected nodes in $c$.
\begin{figure}[!ht]
    \centering
    \includegraphics[width=0.75\linewidth]{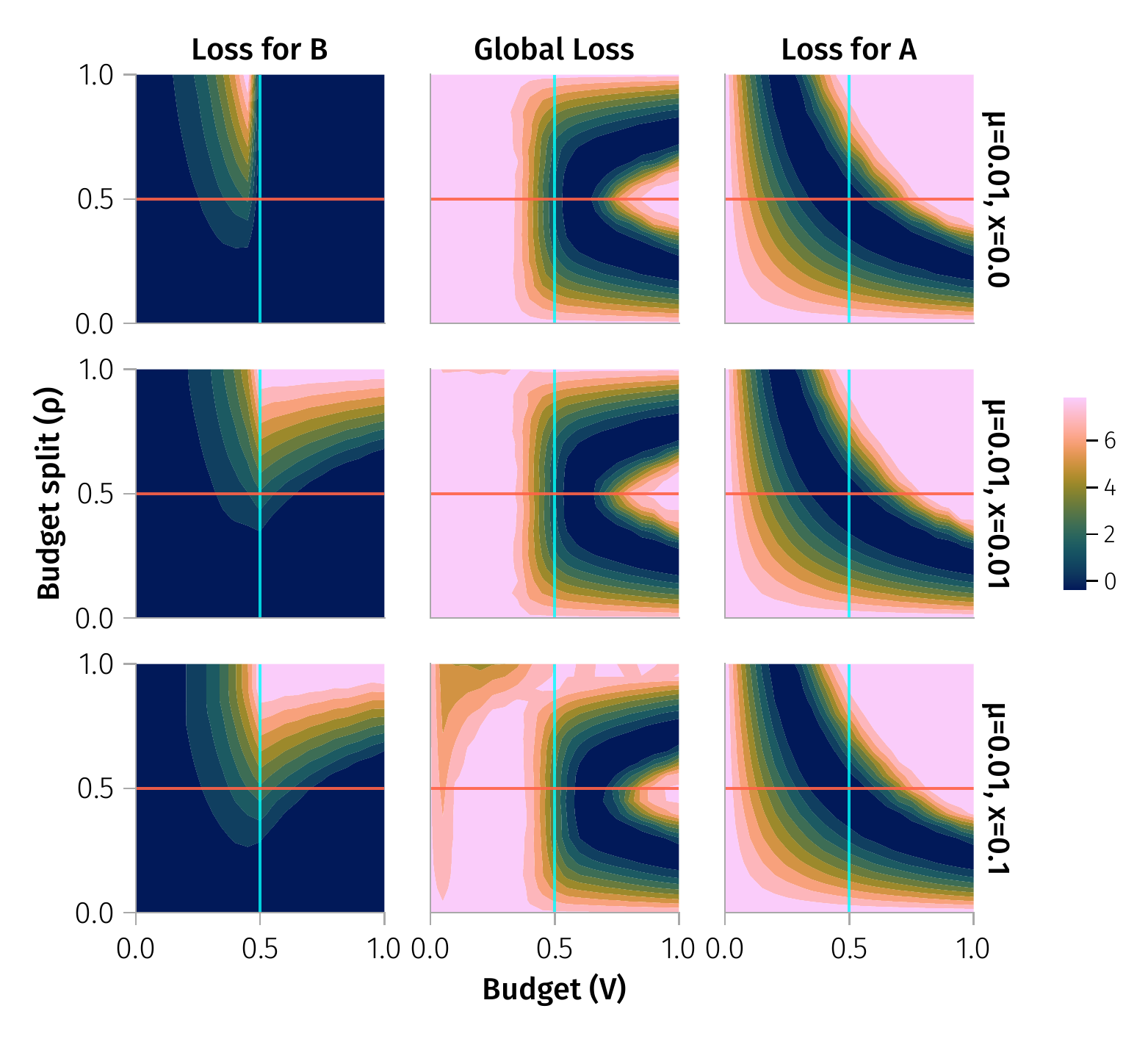}
    \caption{The loss $L^c_{50}$ for all values studied for sharing ratio $\rho$ and vaccine budget $V$ and some $x$ values, the fraction of nodes excluded from the vaccination program. Each column shows the cost for a community $c\in\{B,A+B \text{ (Global)},A\}$.}
    \label{fig:loss}
\end{figure}

Figure~\ref{fig:loss} shows a large region in parameter space for which the cost of sharing is close to zero for the contributing community, even when full vaccination coverage is sacrificed. 
It is clear that sharing may lead to local and global benefits, specially when $B$ has a surplus of vaccine.
The effect of increasing the fraction of nodes in $B$ excluded from vaccination is to reduce the area in parameter space for which sharing is beneficial for $B$.

\begin{figure}[!ht]
    \centering
    \includegraphics[width=0.75\linewidth]{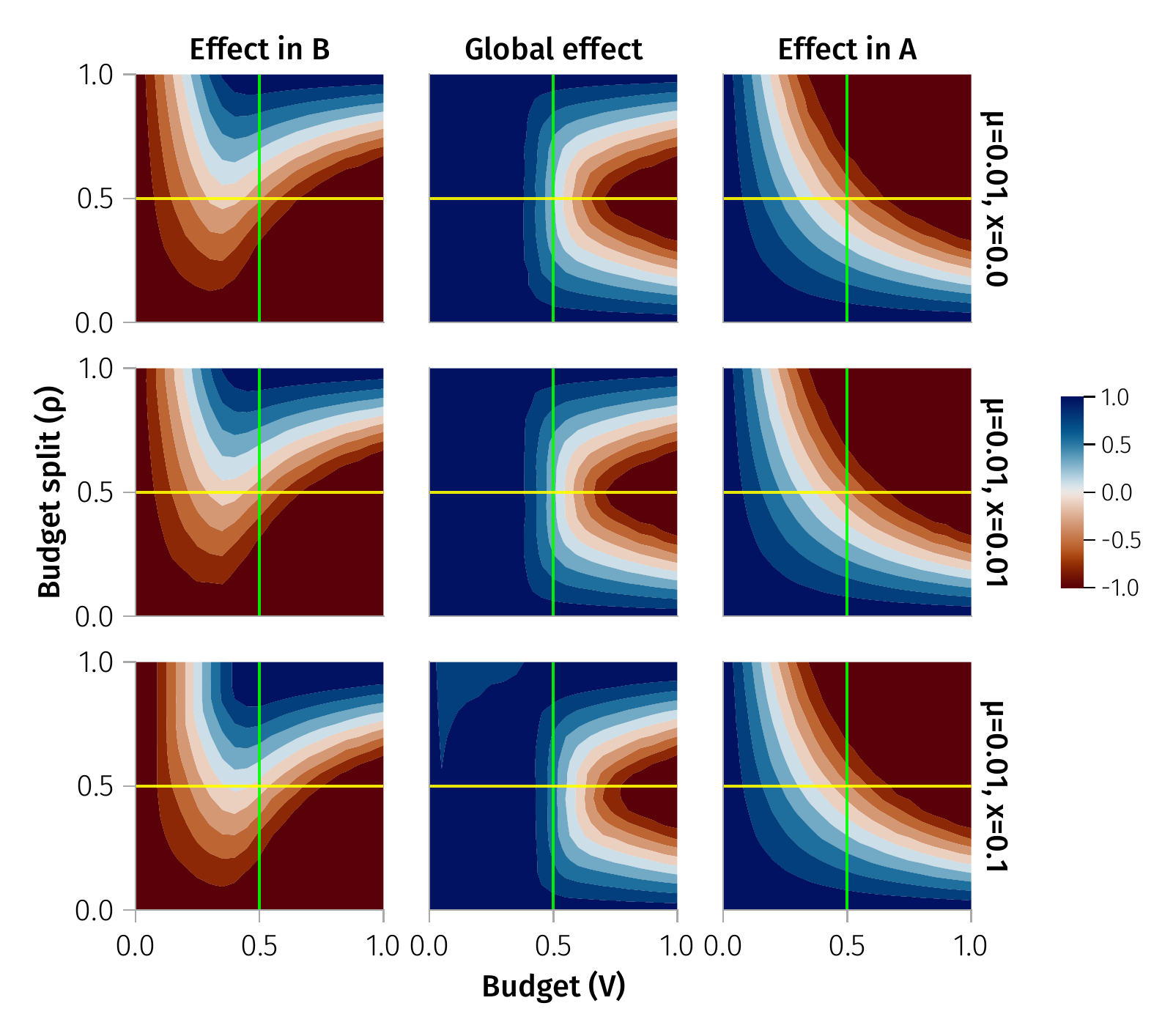}
    \caption{The difference $n^c_t(\rho) - n^c_t(\rho=0)$ between $\rho\neq 0$ and $\rho=0$ in $L^c_t$ for all studied values of budget split $\rho$ and budget $V$ and  some values $x$, the fraction of nodes excluded from vaccination. Columns show the cost for each community $c\in\{B,A+B\text{ (Global)},A\}$, under an strict isolation policy with $\mu=0.01$.}
    \label{fig:diff}
\end{figure}

The difference between a share ratio $\rho\neq 0$ and no sharing, $\rho=0$, in the total fraction of infected nodes $n^c_t(\rho) - n^c_t(\rho=0)$ is shown in Fig.~\ref{fig:diff}


\section{\label{sec:discussion}Discussion}
We investigate the futility of being selfish in vaccine distribution through the application of the SIR model and optimal control for epidemic spreading in a network with community structure.
A vaccine budget $V\in[0,1]$ is split $\rho\in[0,1]$ between the two communities $A$ and $B$, assumed to be under the control of the later. We study the epidemic spread in the parametric space of $V$ and $\rho$, where we look at the fraction of nodes in states $I$ or $R$, except those who have been already vaccinated.
The vaccine budget takes values from $[0,1]$, meaning the fraction of nodes in the whole network, not only $B$, that can be fully immunized, while the share ratio means the fraction of the budget allocated to community $A$.

Employing the DMP framework to estimate the marginal distribution for node states through the epidemic evolution, the model shows a large area in parameter space where $B$ pays little or no cost, in terms of total fraction of infected nodes, for sharing vaccines with $A$, while benefiting the global population.
This result can be viewed as a theoretical complement to~\cite{mooreRetrospectivelyModelingEffects2022}, which retroactively estimates the possible number of lives that could have been saved if a more balanced distribution of vaccines was promoted across the world.
Given the advancements in contact tracing technology, the reconstruction of network topology and the application of optimal control techniques, help to devise a vaccination strategy that shows the futility of selfishly hoarding vaccines. It strengthens the argument for developing a collective solution for a global problem.

\vspace*{0.5cm}
\section*{Acknowledgements}
This work is supported by the Leverhulme trust (RPG-2018-092).
\vspace*{0.5cm}

\appendix
\section{Network topology}
The spreading processes of interest in this work takes place on an interaction graph of two connected components, $A$ and $B$, with the same degree distribution; these are initially disconnected from each other, with $N_A$ and $N_B$ nodes respectively, with a degree distribution given by a power law of degree $\alpha$ and minimum degree $k_0$.
Graphs with power law distribution are called \emph{scale-free} and have some of the properties observed in many real world interaction networks, such as clustering coefficients similar with acquaintance relationships~\cite{newman2018networks,Albert_2002}.
To connect the $A$ and $B$ components, $E_A$ and $E_B$ edges are randomly selected from each component such that $\mu=\frac{E_A + E_B}{E}$, where $E$ is the total number of edges and $\mu$ the fraction of edges connected between communities. Each node still belongs to its initial community, i.e., it has at least as many neighbors within its community as to the other. 
Figure~\ref{fig:example-network} shows an example of such networks.
Controlling the number of edges between components allows for the investigation of how community isolation/interconnection influences vaccine sharing strategies.

\section{\label{appendix-dmp}DMP equations}
The DMP equations for the SIR model of Eq.~\ref{eq:sir-rules} are
\begin{align}
    \theta^{k\to i}(t+1) & = \theta^{k\to i}(t) - \beta^{ki}\phi^{k\to i}(t)\\
    P^{k\to i}_S(t+1) & = P^k_S(0)\prod_{t'=0}^t(1-\nu^k(t'))\prod_{j\in\partial k/i}\theta^{j\to k}(t+1)\\
    \phi^{k\to i}(t+1) & = (1-\beta^{ki})(1-\gamma^{k})\phi^{k\to i}(t)\nonumber \\
     & - [P^{k\to i}_S(t+1) - P^{k\to i}_S(t)]
\end{align}
and for the edge messages and for the node messages,
\begin{align}
    P^i_S(t+1) & = P^i_S(0)\prod_{j\in\partial i}\theta^{j\to i}(t+1)\\
    P^i_R(t+1) & = P^i_R(t) + \gamma^i P^i_I(t)\\
    P^i_I(t+1) & = 1 - P^i_S(t+1) - P^i_R(t+1)
\end{align}

We use the DMP dynamics equations in conjunction with optimal control-based optimization in order to extremize an objective function as described in~\cite{Lokhov_2017}
The objective function to be optimised and the constraints for budget, dynamics and initial conditions give a Lagrangian $\mathcal{L} = \mathcal{O} + \mathcal{B} + \mathcal{D} + \mathcal{I} + \textbf{P}$, where $\mathcal{P}$ is a constraint bounding $\nu^i(t)$ withing a probability interval, $\mathcal{O} = \sum_{i\in B}\sum_{t=0}^T (P_S^i(t)+p_R^i(t)\ln(P_S^i(t)+P_R^i(t))$ is the objective function, $\mathcal{D}$ and $\mathcal{I}$ are constraints enforcing the DMP equations and initial conditions and $\mathcal{B}$ enforces the vaccination budget.
To optimise the objective, we first solve the DMP equations, the forward pass, and then solve the Lagrange multipliers equations backward in time, given by the derivatives of $\mathcal{L}$ with respect to the messages, and budget allocation $\nu$. The calculation is long and omitted here, but it is almost identical to the one presented in~\cite{Lokhov_2017} including the way the equations are solved.

The parameters used for network creation and budget control are presented in Table~\ref{tab:parameters}
\begin{table}[h]
    \centering
    \begin{tabular}{c c}
        Number of nodes: & $N = 1024$  \\
        Number of nodes in community: & $N_A = N_B = 512$ \\
        Power law exponent: & $\alpha = 3.1$ \\
        Minimum node degree: & $k_0 = 3$ \\
        Fraction of edges between $A$ and $B$: & $\mu \in \{0.001,0.01,0.1\}$ \\
        Fraction of nodes covered by vaccine: & $V \in [0,1]$\\
        Split ratio of $V$ between $A$ and $B$: & $\rho \in [0,1]$\\
        Fraction of nodes excluded from $V$: & $x \in \{0,0.01,0.1\}$
    \end{tabular}
    \caption{Parameter values}
    \label{tab:parameters}
\end{table}

The initial condition used for the DMP equations in the preemptive sharing scenario (no optimal control) are
\begin{align} \label{eq:initial-values}
\begin{split}
    P^a_R(0) & = \frac{\rho V N}{N_A}(1-x)\\
    P^b_R(0) & = \frac{(1-\rho) V N}{N_A}(1-x) \\
    P^a_I(0) & = \frac{1-P^a_R(0)}{N_A} \\
    P^b_I(0) & = \frac{1-P^b_R(0)}{N_B} \\
    P^i_S(0) & = 1 - P^i_I(0) - P^i_R(0) \\
    \theta^{k\to i}(0) & = 1\\
    \phi^{k\to i}(0) & = P^k_I(0)\\
    P^{k\to i}_S(0) & = P^k_S(0)\\
\end{split}
\end{align}
while $P_R^a(0)=P_R^b(0)=0$ for the optimal control approach.

\bibliographystyle{unsrt}
\bibliography{tfobrefs}

\end{document}